\documentclass[pra,twocolumn,aps,amsmath,showkeys,showpacs,floatfix,nofootinbib]{revtex4}
\usepackage{graphicx}
\usepackage{dcolumn}
\usepackage{bm}
\usepackage{color}

\def\beq{\begin{equation}}
\def\eeq{\end{equation}}
\def\beqa{\begin{eqnarray}}
\def\eeqa{\end{eqnarray}}
\begin{document}


\title{Shortcut to adiabaticity in internal bosonic Josephson junctions}

\author{A. Yuste$^1$, B. Juli\'a-D\'\i az$^{1,2}$, E. Torrontegui$^3$, 
J. Martorell$^1$, J. G. Muga$^{3,4}$, and A. Polls$^1$}

\affiliation{$^1$Deptartament d'Estructura i Constituents de la Mat\`eria, 
Facultat de F\'isica, U. Barcelona, 08028 Barcelona, Spain}
\affiliation{$^2$ICFO-Institut de Ci\`encies Fot\`oniques, 
Parc Mediterrani de la Tecnologia, 08860 Barcelona, Spain}
\affiliation{$^3$ Departamento de Qu\'\i mica-F\'\i sica, 
UPV-EHU, Apartado 644, 48080 Bilbao, Spain}
\affiliation{$^4$ Department of Physics, Shanghai University, 
200444 Shanghai, People's Republic of China}

\begin{abstract}
We extend a recent method 
to shortcut the adiabatic following to internal bosonic Josephson 
junctions in which the control parameter is the linear coupling 
between the modes. The approach is based on the mapping between the two-site 
Bose-Hubbard Hamiltonian and a 1D effective Schr\"odinger-like 
equation, valid in the large $N$ (number of particles) limit. 
Our method can be readily implemented in current internal bosonic 
Josephson junctions and it improves substantially the production 
of spin-squeezing with respect to usually employed linear rampings. 
\end{abstract}

\maketitle

\section{Introduction}

Practical applications of quantum technologies will require 
the controlled production of many-body correlated quantum 
states, in particular ground states (g.s.). It is thus desirable 
to find efficient mechanisms for their fast production. 
Bosonic Josephson junctions (BJJs) are among the simplest 
systems whose ground states already contain many-body 
correlations beyond mean field. Schematically, BJJs  are 
ultracold bosonic vapors in which, to a good approximation,  
the atoms populate only two mutually interacting single-particle 
levels. Recently, BJJs have been studied experimentally by 
several groups~\cite{Albiez05,esteve08,gross10,riedel10,zib10,amo13,berrada13}. 
Current nomenclature calls external Josephson junctions those 
in which the two levels are spatially separated, usually 
by means of a potential barrier ~\cite{Albiez05,GO07,riedel10,berrada13}. 
In internal Josephson junctions instead, the two levels are 
internal to the same atom~\cite{zib10}. The two-site Bose-Hubbard 
Hamiltonian provides a suitable theoretical description of 
both internal and external junctions~\cite{lipkin,Mil97,Leggett01,GO07}. 
A notable feature of this simple Hamiltonian is that, within 
subspaces of fixed number of particles, it can be mapped into 
an SU(2) spin model. This makes these systems suitable  to 
study  very squeezed spin states~\cite{wine92,kita}, as 
proven experimentally in Refs.~\cite{esteve08,riedel10}.

In previous work~\cite{oursmuga} we described how a method to 
shortcut the adiabatic following in elementary quantum mechanical 
systems could be applied to produce of spin-squeezed states 
in BJJs. In particular, we adapted a simple method developed 
for harmonic oscillators in which the frequency could be varied 
in time~\cite{muga1,muga2}. In ~\cite{oursmuga} we described 
the most straightforward application, where
the inter-atomic interaction strength was the control parameter. This 
is nowadays a parameter that can be varied experimentally but 
it is difficult to control with good accuracy on the time 
scales considered. To overcome this problem, here we will extend 
the earlier protocol by varying instead the linear coupling between 
the states (atomic levels) in internal junctions~\footnote{In 
external junctions this can be done by increasing the barrier 
height between the two wells}. This variation can be done with 
fantastic accuracy~\cite{zib10,ziboldphd,nicklasphd} and we shall 
focus on this case.

The protocols to shortcut adiabatic evolution are generally designed 
to drive in a finite time a system from some initial state to a final 
state that could be reached adiabatically. An important advantage of 
these protocols is that they can, in addition, aim at controlling 
other properties during the evolution, e.g. reducing transient energy 
excitation, noise sensitivity or optimizing other quantities of 
interest~\cite{muga4,njp12,pra13}. In addition, formulas to achieve 
shortcuts to  adiabatic following  are analytic for harmonic 
oscillator Hamiltonians~\cite{muga1}. From the experimental point 
of view the methods are capable to produce a stationary eigenstate 
of the Hamiltonian at the final time, making it unnecessary to 
stop or freeze the dynamics. 

The paper is organized as follows. First in Section~\ref{sec:tf} 
we describe the theoretical framework. In Sec.~\ref{sec:num} we 
present our numerical results, including a specific subsection 
with parameter values within reach with current experiments. 
Finally, in Sec.~\ref{sec:sum} we summarize the results and 
provide some concluding remarks.

\section{Theoretical framework}
\label{sec:tf}

The dynamics of a BJJ  can be well described by a quantized 
two-mode model~\cite{Mil97,Leggett01,GO07}, the Bose Hubbard 
Hamiltonian $H= \hbar {\cal H}_{\rm BH}$, 
\beq
{\cal H}_{\rm BH}= -2J \hat{J}_x + U \hat{J}_z^2  \,,
\label{eq:bh}
\eeq
where the pseudo-angular momentum operator 
$\hat{{\bf J}}\equiv \{\hat{J}_x,\hat{J}_y,\hat{J}_z\}$ 
is defined as
\beqa
{\hat J}_x &=& {1\over 2} (\hat{a}_1^{\dag} \hat{a}_2 
+ \hat{a}_2^{\dag} \hat{a}_1),\nonumber\\
{\hat J}_y &=& {1\over 2i}(\hat{a}_1^{\dag} \hat{a}_2 
- \hat{a}_2^{\dag} \hat{a}_1),\nonumber\\
{\hat J}_z &=& {1\over 2} (\hat{a}_1^{\dag} \hat{a}_1 
- \hat{a}_2^{\dag} \hat{a}_2),
\eeqa
and $\hat{a}_j^{\dag} (\hat{a}_j) $ creates (annihilates) a boson 
at site $j$. For bosons: $[\hat{a}_i,\hat{a}_j^{\dag}] =\delta_{i,j}$. 
$J$ is the hopping strength, and $U$ is the non-linear coupling 
strength proportional to the atom-atom $s$-wave scattering length. 
In internal BJJs, $U$ is proportional to $a_{1,1}+a_{2,2}-2 a_{1,2}$, 
with $a_{1,1}$ and $a_{2,2}$ the intra-species scattering lengths and 
$a_{1,2}$ the inter-species one~\cite{zib10}. In this work we consider 
repulsive interactions, $U>0$. For internal BJJs, the inter-species 
$s$-wave scattering length in $^{87}$Rb atoms can be varied by applying 
an external magnetic field thanks to a well characterized Feshbach 
resonance at $B =9.1$ G, as done in Ref.~\cite{zib10} for the setup 
that we are considering. In this work instead, we assume a time dependent
hopping strength, $J(t)$, keeping $U$ and $N$ fixed during the time 
evolution, which should be simpler and more accurate from 
an experimental point of view. 

The time dependent Schr\"odinger equation (TDSE) is written as
\beq
\imath  \partial_t |\Psi\rangle = {\cal H}_{\rm BH} |\Psi\rangle\,.
\eeq
For a given $N$, an appropriate many-body basis 
is the Fock basis, $\{ | m_z = (N_1-N_2)/2 \rangle\}$, with 
$m_z=-N/2,\dots, N/2$. A general many-body state, 
$|\Psi\rangle$, can then be written as 
\beq
|\Psi\rangle = \sum_{m_z=-N/2}^{N/2} c_{m_z} |m_z\rangle \,.
\label{eq:wf}
\eeq 
It is useful to define the population imbalance of the state 
as $z \equiv m_z/(N/2)$. 

For a given state, the Kitagawa-Ueda spin squeezing parameter~\cite{kita}, 
termed also number squeezing parameter~\cite{esteve08}, is defined as
$\xi_N^2(t)=\Delta \hat{J}_z^2 / (\Delta \hat{J}_z^2 )_{\rm ref}$, 
where 
$\Delta \hat{J}_z^2 \equiv \langle {\hat J}_z^2\rangle 
- \langle {\hat J}_z\rangle^2$ and 
$(\Delta \hat{J}_z^2 )_{\rm ref}=N/4$ is the value for a coherent
state with $\langle \hat{J}_y\rangle=\langle \hat{J}_z\rangle=0$. 
The many-body state is said to be number-squeezed when 
$\xi_N<1$~\cite{kita}. The Wineland spin-squeezing parameter~\cite{wine92}, 
also referred to as coherent spin-squeezing parameter~\cite{gross10}, is 
defined as~\cite{wine92,Sorensen2001}
$
\xi_S^2=  N (\Delta \hat{J}_z^2) /
\langle \hat{J}_x\rangle^2 =  
\xi_N^2 / \alpha^2,
$
where the phase coherence of the many-body state is 
$
\alpha(t) = \langle \Psi(t)| 2\hat{J}_x/N|\Psi(t) \rangle$.
$\xi_S$ takes into account the delicate compromise 
between improvements in number-squeezing and loss of 
coherence. States with $\xi_S<1$ have been proposed to 
be the basis of a new Ramsey type atom interferometer with 
increased phase precision (compared to that of the coherent 
spin states). This gain in precision can be directly 
related to entanglement in the system~\cite{Sorensen2001}.

Since we take $J$ as the control parameter we slightly 
detour from the derivation in Refs.~\cite{ST08,ours10-2,oursober}. 
Following similar steps as described in those references,  
one can obtain in the semiclassical ${\eta}\equiv 1/N \ll 1$ limit 
a Schr\"odinger-like equation 
\begin{equation}
i \eta  \partial_t \psi(z,t) = {\cal H}_N \ \psi(z,t)
\label{eq:tdse} 
\end{equation}
for the continuous extrapolation 
of $z$, where 
\begin{eqnarray}
{\cal H}_N(z) \psi(z)  
&\equiv&-2 \eta^2 J \ \partial_z \sqrt{1-z^2}\partial_z \psi(z) 
+ {\cal V}(z) \psi(z),\nonumber \\
\label{eq:pse}
\end{eqnarray}
and  ${\cal H}_N \equiv \eta {\cal H}_{\rm BH}$, 
${\cal V}(z)= -J \sqrt{1-z^2}+(1/2) (NU/2) z^2$.  
$\psi(z) = \sqrt{N/2}\, c_{m_z}$ is normalized as 
$\int_{-1}^1 dz |\psi(z)|^2=1$. 
\begin{figure}
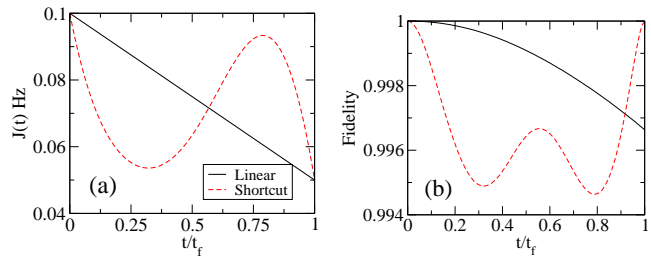

\vspace{10pt}
\includegraphics[width=41mm, angle=0]{jt1.eps}
\hspace{2pt}
\includegraphics[width=41mm, angle=0]{jolap1.eps}
\caption{(a) $J(t)$ used in the shortcut protocol compared to 
the corresponding linear ramping. The initial and final values 
of $\gamma$ are $\gamma_i$=10, $\gamma_f=$20, and 
$t_f=0.08$ $t_{\rm Rabi}^{(i)}$. In panel (b) we depict the fidelity 
(overlap) between the evolved state and the instantaneous ground 
state. The number of particles and non-linearity 
are, $N=100$, and $U=1/(50 t_{\rm Rabi}^{(i)})$ , respectively.}
\label{j1}
\end{figure}

\begin{figure*}[t]
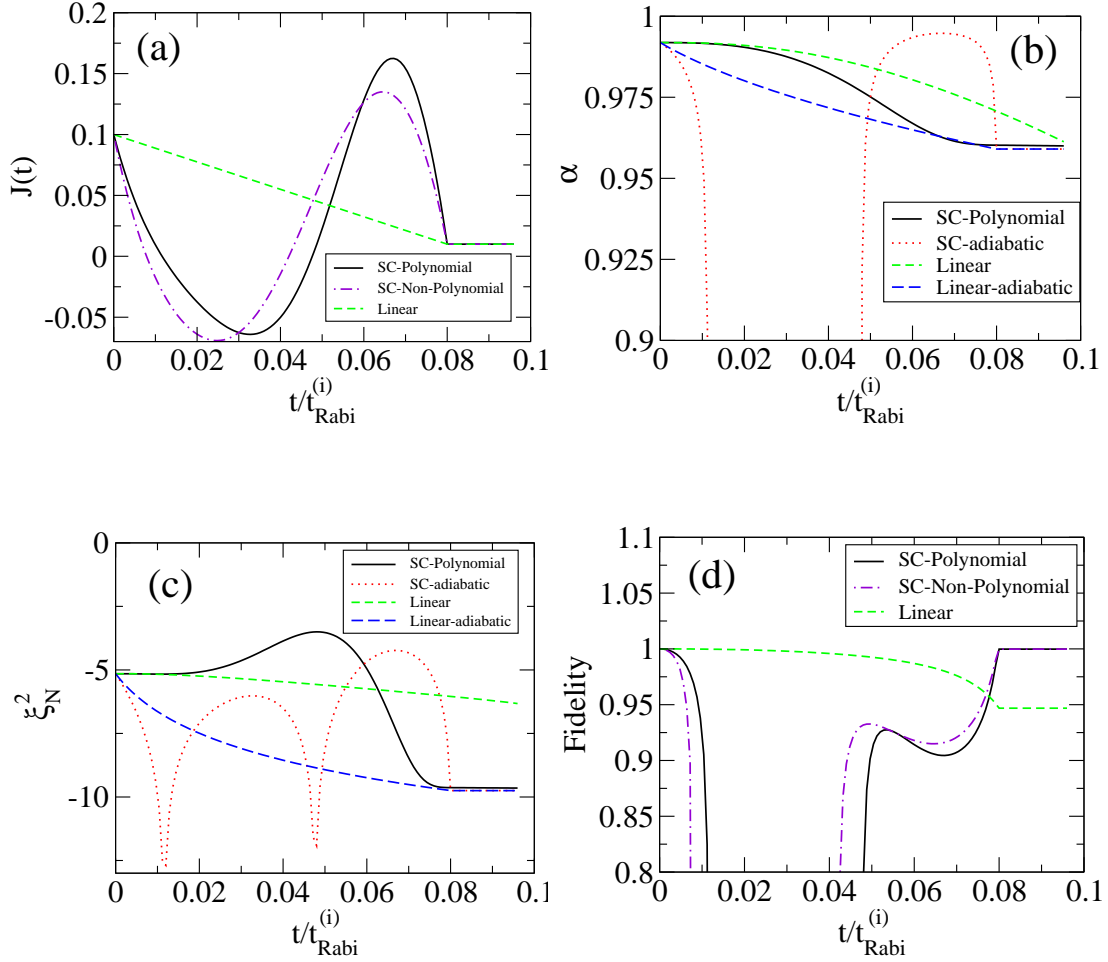

\vspace{10pt}
\includegraphics[width=72mm, angle=0]{j2.eps}
\includegraphics[width=72mm, angle=0]{jcohe2.eps}
\vspace{37pt}

\includegraphics[width=72mm, angle=0]{jsque2.eps}
\includegraphics[width=72mm, angle=0]{jolap2.eps}
\caption{Panel (a): evolution of $J(t)$, (b) coherence of the 
state, (c) its number squeezing,  and (d) the instantaneous 
fidelity. $N=100$ atoms, $\gamma_i=10$, $\gamma_f=100$ and 
$t_f=0.08$ $t_{\rm Rabi}^{(i)}$. For $t>t_f$ we fix $\gamma(t)=\gamma(t_f)$.  
\label{j2}}
\end{figure*}

For repulsive atom-atom interactions the potential in Fock space, 
${\cal V}(z)$, is to a very good approximation a harmonic oscillator.
Neglecting the $z$ dependence of the effective 
mass term and expanding $\sqrt{1-z^2} \simeq 1- z^2/2$ 
in ${\cal V}(z)$, the Hamiltonian in Eq.~(\ref{eq:pse}) reduces to 
\beq
{\cal H}_N \simeq -2 J \eta^2 \partial_z^2 +  \frac{1}{2} (J+NU/2) z^2, 
\label{eq:parab}
\eeq 
A difference with respect to Ref.~\cite{oursmuga} and to previous 
applications of  shortcuts-to-adiabaticity to harmonic-oscillator 
expansions is that now the control parameter $J(t)$ shows up both 
as a formal time-dependent (inverse of) mass and as an additive term in 
the force constant. In the Appendix ~\ref{app:ex} we 
provide the  extension of the shortcut technique for  
this type of time dependence when $NU/2\gg |J|$, so that we can 
approximate $(J+NU/2)\simeq {NU/2}$. 
Defining $\gamma=NU/(2J)$, this limit corresponds to $\gamma\gg1$, which 
is easily attainable in current experiments. It is also relevant as 
it corresponds to very spin-squeezed ground states 
of the bosonic Josephson junction. In Appendix~\ref{app:bd} we verify 
that the method is not applicable when $|\gamma|<1$. 

For the case at hand, the inverse engineering described 
in~Appendix~\ref{app:ex} 
translates into solving for $J(t)$ in the following Ermakov equation,
\begin{equation}
{\ddot b} -2\frac{({\dot b})^2}{b} =  \frac{4 k}{J(t)} b 
- \frac{k^2}{\eta^2} b^5,
\label{eq:erm}
\end{equation}
where the dots indicate time derivatives,  and $k=NU/2$ and $b(t)$ must satisfy the boundary conditions 
\begin{eqnarray}
b_0 &\equiv& b(0)= \left( {8 \eta^2 J(0)\over N U}\right)^{1/4},
\label{boco}\\
b_f &\equiv& b(t_f)= \left( {8 \eta^2 J(t_f)\over N U}\right)^{1/4}, 
\nonumber\\
\dot{b}(0)&=& \ddot{b}(0)=\dot{b}(t_f)=\ddot{b}(t_f)=0.
\end{eqnarray}
For simplicity we apply  
the polynomial~\cite{muga1}
\beqa
b_{\rm poly}(t) &=&b_0
+10(b_f-b_0)s^3
-15(b_f-b_0)s^4 \nonumber\\
&&+6(b_f-b_0)s^5.
\eeqa
with $s=t/t_f$. We also consider a non-polynomial form 
in some comparisons, 
\beq
b_{\rm non-poly}(t) = b_0 \left({b_{f} \over b_{0}}\right)^{6s^5-15s^4+10s^3}.
\eeq

\begin{figure*}[t]
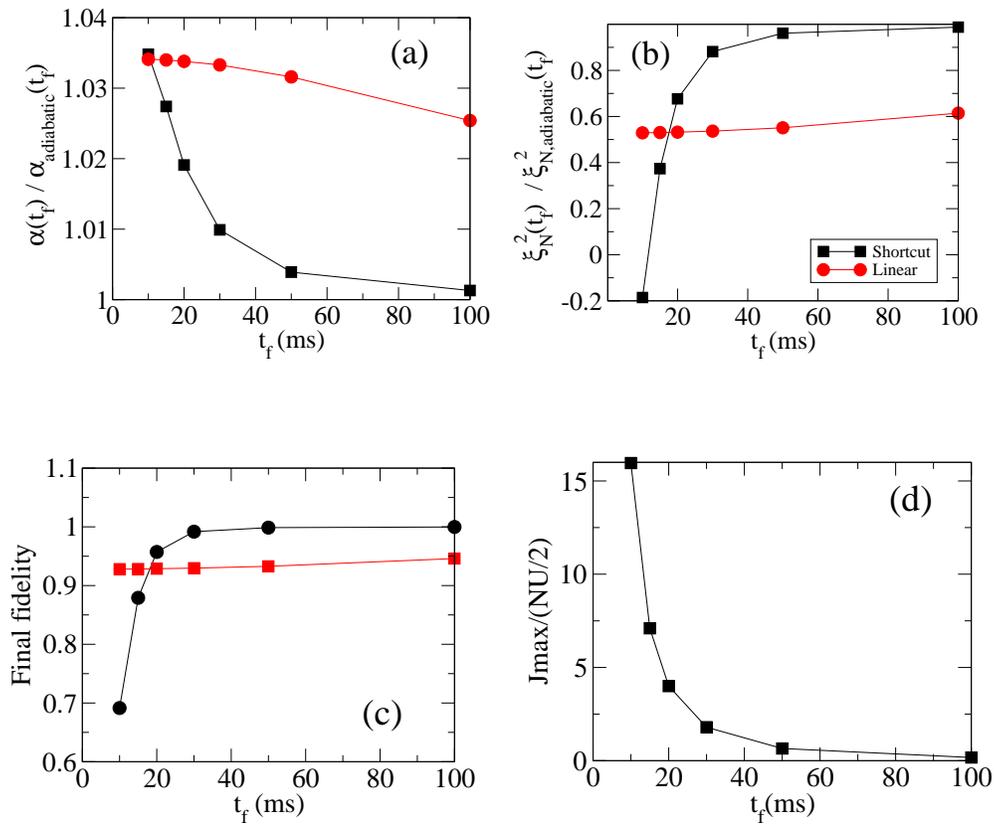

\vspace{35pt}

\includegraphics[width=62mm, angle=0]{jcoh7.eps}
\hspace{2pt}
\includegraphics[width=62mm, angle=0]{jsque7.eps}

\vspace{35pt}

\includegraphics[width=62mm, angle=0]{jolap7.eps}
\hspace{13pt}
\includegraphics[width=62mm, angle=0]{maxj.eps}
\caption{Properties of the state of the system at $t_f$ after 
a shortcut protocol (black) and a linear ramp (red), for 
different values of $t_f$. (a) depicts the relative coherence, 
$\alpha/\alpha_{\rm adiabatic}$. (b) shows the relative number 
squeezing, $\xi^{2}_{N}/\xi^{2}_{N, \rm adiabatic}$. (c) contains the 
value of the final fidelity. (d) shows the maximum value of 
$1/\gamma$ required for the shortcut process. In all simulations
$\gamma_i=10$, $\gamma_f=100$, $N=100$, and $U=0.49$ Hz.}
\label{j7}
\end{figure*}

\section{Numerical simulations of the shortcut protocol}
\label{sec:num}

In all cases we will consider the evolution from an initial 
g.s. corresponding to $\gamma = \gamma_i$ to a final one with 
$\gamma= \gamma_f$. The control parameter $J(t)$, will go 
from $J(0)=J_i$ to $J(t_f)=J_f$ in a time $t_f$ with a fixed value 
of $U$. In our first application, we will measure the time in units 
of the initial Rabi time, $t_{\rm Rabi}^{(i)} =\pi/J_i$. Later, 
we will consider realistic values of $U$ and $t$ taken from 
recent experiments. 

In Fig.~\ref{j1} we consider a factor 2 change in $\gamma$, 
from  $\gamma_i=10$ to $\gamma_f=20$ in a time 
$t_f=0.08$ $t_{\rm Rabi}^{(i)}$, with $N=100$ and $NU/2=1/t_{\rm Rabi}^{(i)}$. 
We compare the shortcut protocol using the polynomial 
ansatz for $b(t)$ to a linear ramping: $J(t)=J_i  + (J_f-J_i)(t/t_f)$. 
The shortcut method is shown to work almost perfectly, and we obtain 
a final fidelity $\simeq 1$ (despite the process being diabatic 
during the intermediate evolution). For this case, the linear 
ramping produces a final fidelity of $\simeq 0.9965$. As 
it occurred with the harmonic oscillator~\cite{muga1} or in  
Ref.~\cite{oursmuga}, for more stringent processes, i.e. shorter 
final times or larger changes in $\gamma$, the method requires 
negative values of the control parameter. For instance, if we 
require a factor of 10 change, from  $\gamma_i=10 $ to 
$\gamma_f=100$ under the same conditions, $J(t)$ becomes 
negative during part of the evolution. Although for usual 
tunneling phenomena the hopping term is always positive, e.g. 
in external Josephson junctions, there are several proposals 
to implement negative or even complex hopping 
terms in optical lattices~\cite{jz03,andre05}. For the 
internal Josephson junctions achieved in Oberthaler's group 
negative tunneling presents no obstacle as they are able of 
engineering a tunneling term of the form (see Sect. 3.5 of 
Ref.~\cite{ziboldphd})
\beq
J(t) \left[ \hat{J}_x \ \cos\phi_c(t)) + \hat{J_y} \ \sin\phi_c(t)\right]
\eeq
with $\phi_c(t)$ a phase which can be controlled externally. 

Our results are shown in Fig.~\ref{j2}. First we see that, 
for both polynomial 
and non-polynomial choices of $b(t)$ described above, $J$ 
changes its sign at intermediate times, see  Fig.~\ref{j2}(a). 
This implies a transient loss of fidelity (overlap) between 
the evolved state and the instantaneous ground state of the 
system, as shown in  Fig.~\ref{j2}(d). With the shortcut protocol 
both the coherence,  Fig.~\ref{j2}(b), and number squeezing,  Fig.~\ref{j2}(c), 
evolve smoothly towards their adiabatic value. In contrast, the 
linear ramping fails to provide the adiabatic values at the 
final time. The instantaneous ground state coherence, dotted red
line in  Fig.~\ref{j2}(b), is rather involved as it follows the $J(t)$ path. 
As seen in  Fig.~\ref{j2}(c) the linear squeezing is $\simeq -6$ dB while the 
adiabatic one, accurately reproduced by the shortcut protocol, is 
$\simeq -10$ dB. This is a notable feature which should be 
experimentally accessible. The linear ramp gives a final fidelity 
of 0.95, well bellow those of the polynomial and non-polynomial 
shortcut protocols which get final fidelities of nearly 1. It is 
also worth stressing that the many-body state produced by the 
shortcut method at $t=t_f$ is almost an eigenstate of the system, 
which implies constant coherence and 
squeezing for $t>t_f$, see Fig.~\ref{j2}(b,c).

\begin{figure*}[t]
\includegraphics[width=60mm, angle=270]{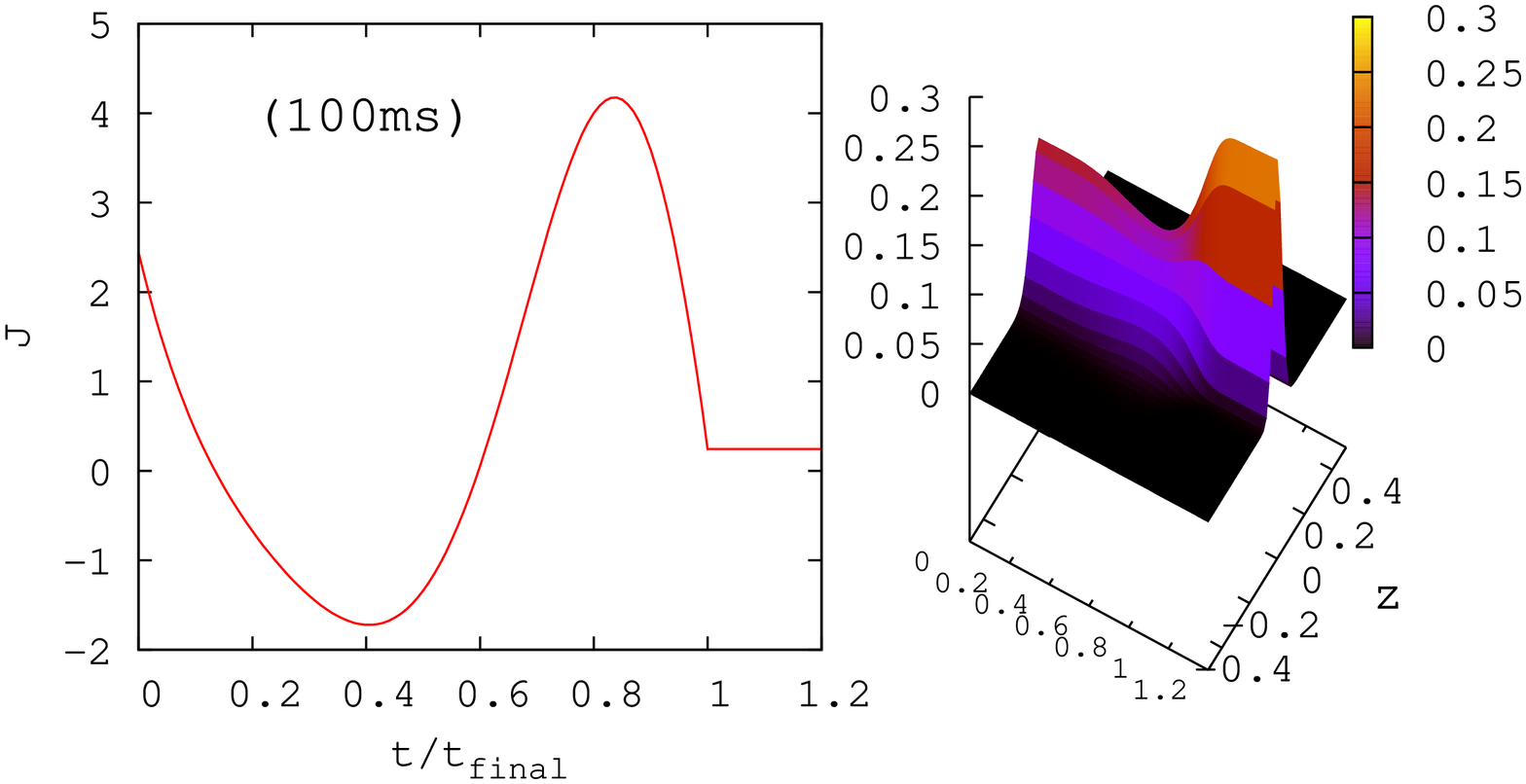}
\includegraphics[width=60mm, angle=270]{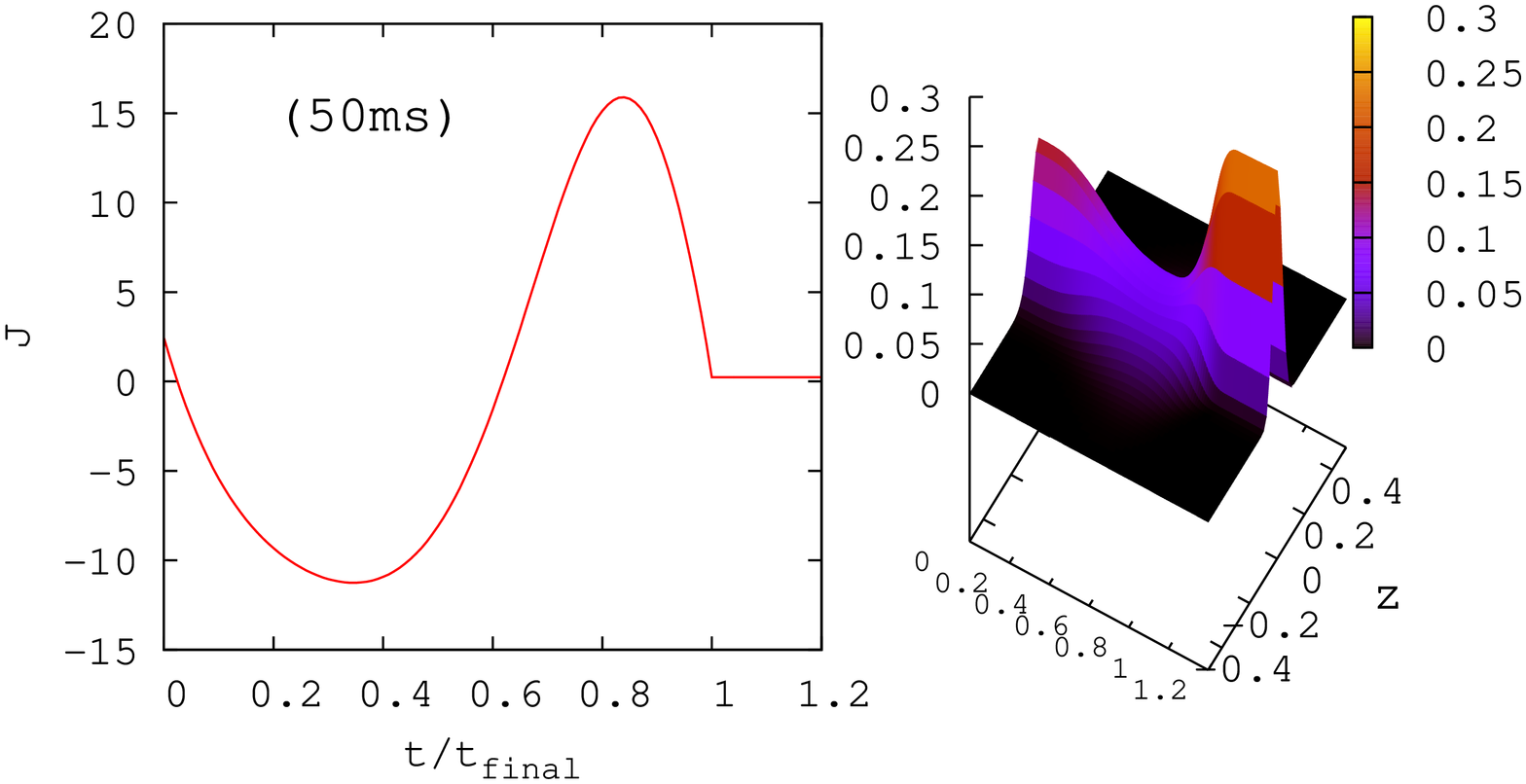}
\includegraphics[width=60mm, angle=270]{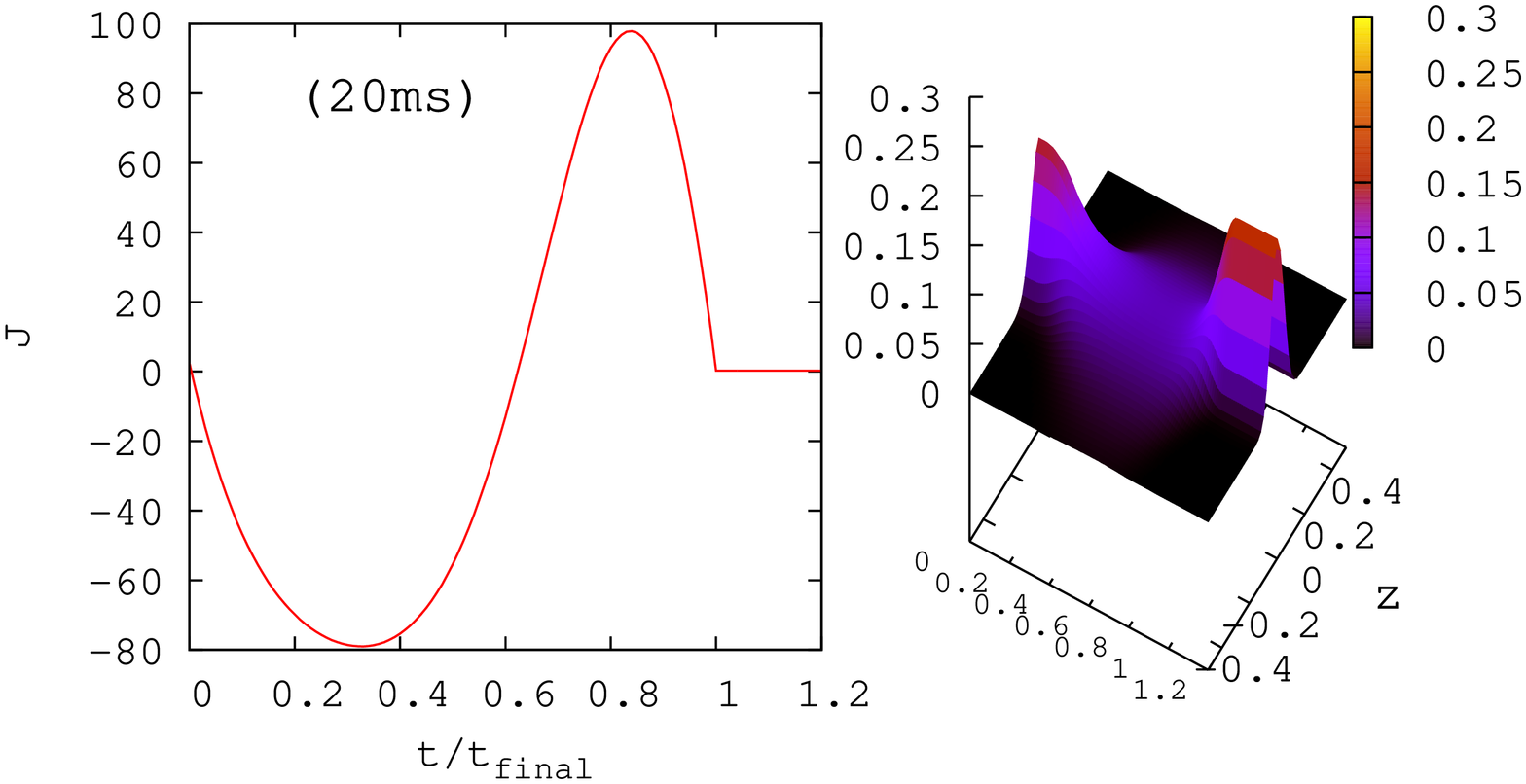}
\includegraphics[width=60mm, angle=270]{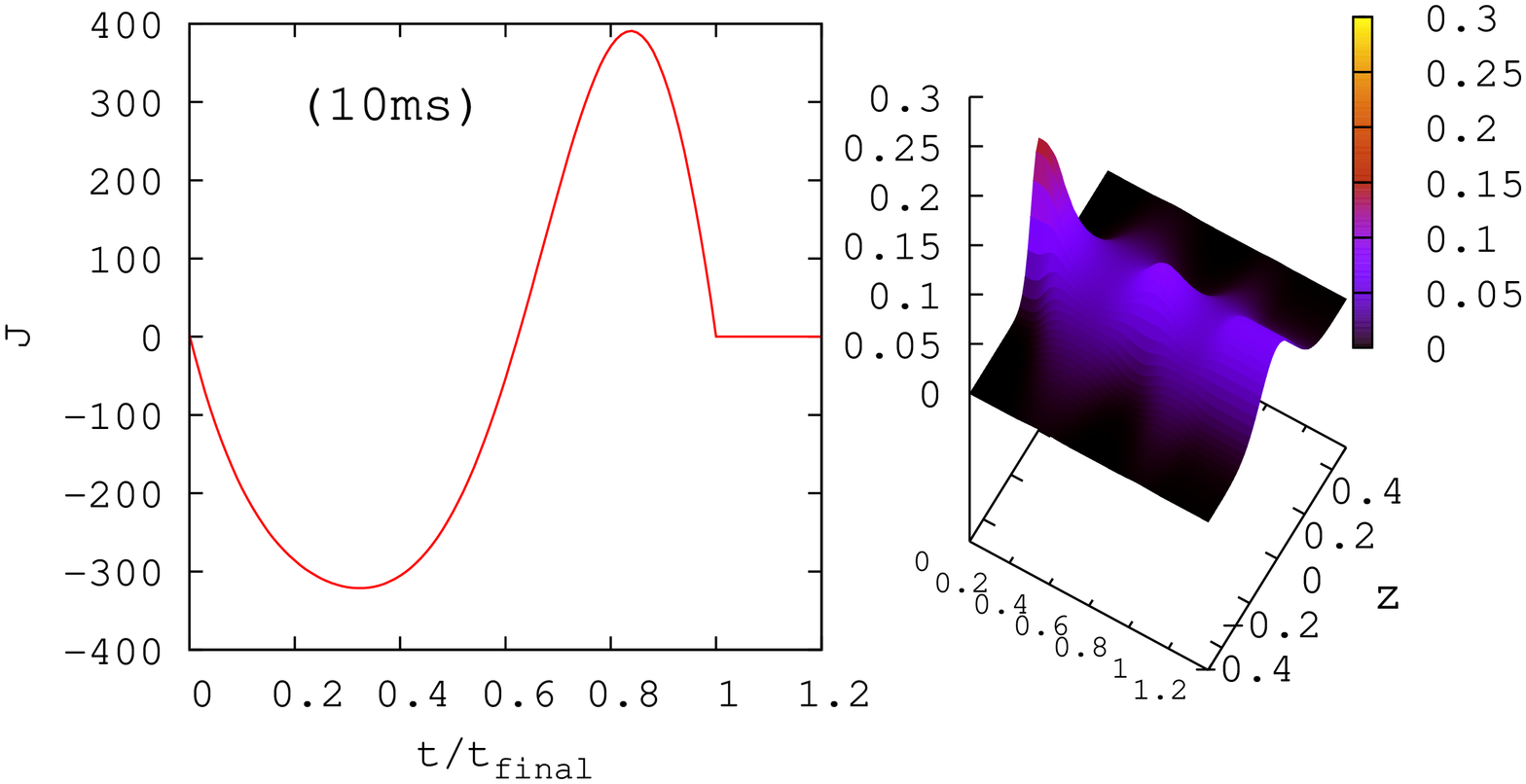}
\caption{(left) Evolution of $J(t)$ required by the polynomial 
shortcut protocol for different final times $t_f=100, 50, 20, 10$ ms. 
(right) Spectral decomposition of the many-body state $|c_z|^2$ 
for the same $t_f$ as a function of time. In all simulations
$\gamma_i=10$, $\gamma_f=100$, $N=100$, and $U=0.49$ Hz.}
\label{wave}
\end{figure*}

It is also possible to engineer fidelity-one processes where 
the control $J(t)$ is constrained from below and above by 
predetermined  values (in particular we could make both 
bounds positive). Prominent examples are the bang-bang 
protocols, with step-wise constant $J$, which solve the 
time-minimization variational problem for given bounds
and boundary conditions \cite{muga2,muga4,salamon,stef10}.

\subsection{Simulations using experimental values for the parameters}
\label{exp}

As explained above, the variation of $J$ with time can be readily 
implemented experimentally. In this section, we will consider 
realistic values of the parameters. Following 
Refs.~\cite{ziboldphd,riedel10} we take a value of the 
non-linearity $U=0.49$ Hz, with $N= 100$ atoms, and make 
variations of $\gamma$ during typical experimental values of 
time: $t_f=10, 15, 30, 50$ and 100 ms. At $t=t_f$ we fix 
$\gamma(t)=\gamma_f$ and evolve the system during an additional 
small time to check whether the state remains close to desired 
final the ground state or not.  

\begin{figure}
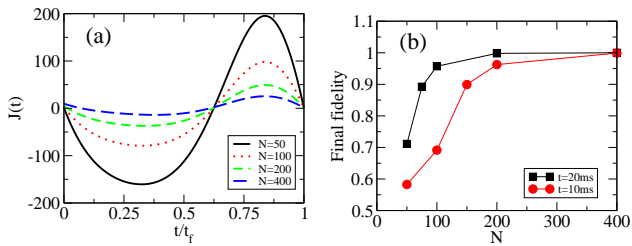

\begin{center}
\includegraphics[width=40mm, angle=0]{jt.eps}
\hspace{2pt}
\includegraphics[width=40mm, angle=0]{atolap.eps}
\caption{(a) Evolution of $J(t)$ required by the polynomial 
shortcut protocol for different values of $N$ and $t_f$=20 ms. 
In (b) we depict the final fidelities attained in the process 
for  $t_f$=10 and $20$ ms. $U=0.49$ Hz and  $\gamma_i=10$, $\gamma_f=100$.}
\label{j9}
\end{center}
\end{figure}

In Fig.~\ref{j7} we depict the final value of the fidelity (c), 
number squeezing (b) and 
coherence of the many-body state (a), as a function of the final 
time imposed $t_f$. The shortcut method (with polynomial ansatz) 
is compared to the linear ramping. The first observation is that 
for $t_f>40$ ms, the shortcut protocol produces a fidelity 
$\simeq 1$, while the linear ramp stays always below $0.95$, 
see (c). Similarly, for $t_f>40$ ms, the final coherence and 
number squeezing are essentially those of the corresponding 
ground state (a,b). This is an important finding, as for 
instance the linear ramping produces roughly half of the number 
squeezing as compared to the adiabatic or shortcut protocol. 
For $t_f<40$ ms, the shortcut protocol is seen to fail, and 
in particular, the achieved final fidelities drop to $0.7$ 
for $t_f=10$ ms, smaller than the linear ramping ones. 
As explained above, our shortcut protocol has been derived 
assuming the validity of a parabolic approximation for the 
potential in Fock space. Therefore we expect the method to 
fail when the intermediate wave 
packet spreads far from the central region in Fock space. In 
Fig.~\ref{wave}, we have plotted the spectral decomposition of 
the many-body state $|c_z|^2$ as a function of time for the same final 
times $t_f$ as above. When the final time is large, the process is 
smooth and the wave function does not spread considerably. 
When we use shorter final times $J(t)$ takes large 
values (so $\gamma$ is small at intermediate times), and the 
effective wave-function spreads considerably in $z$ space. 
A parameter that affects the $J(t)$ functional form is the number 
of atoms $N$. The larger $N$, the smoother the $J(t)$ 
path and the better are the results obtained. 
This is seen in Fig.~\ref{j9}, where we choose only 
two values of $t_f$: $10$ and $20$ ms, and 
consider $N=50, 100, 150$ and $400$ atoms. We also depict 
$J(t)$, which is on average smaller for larger $N$. 

\section{Summary and conclusions}
\label{sec:sum}

We have presented a method to produce ground states of 
bosonic Josephson junctions for repulsive atom-atom 
interactions using protocols to shortcut the adiabatic 
following. We inverse-engineer the accurately 
controllable linear coupling $J$ by mapping a 
Schr\"odinger-like  equation for the (imbalance) 
wavefunction of the Josephson junction onto an ordinary 
harmonic oscillator for which shortcut protocols can be set easily. 
The original equation is a priori more involved for that 
end, as the kinetic-like term includes a time-dependent 
formal mass. As detailed in Appendix ~\ref{app:ex}, the 
mapping requires a reinterpretation of kinetic and potential 
terms, which interchange their roles in a representation 
conjugate to the imbalance. The time dependence of the formal 
mass of the original equation (inversely proportional to $J$) 
implies the time dependence of the frequency of the ordinary 
(constant mass) harmonic oscillator, and $J$ plays finally 
the role of the squared frequency. This mapping is different 
and should be distinguished from the ones used to treat 
harmonic systems with a time dependent mass both in the 
kinetic and the potential terms~\cite{l1}.
From the experimental point of view, our protocol should help 
the production of spin-squeezed states, increasing the maximum 
squeezing attainable in short times. In particular, an important 
shortcoming of recent experimental setups~\cite{ziboldphd}, is that 
they have sizable particle loss on time scales of the order of 
$\simeq 50$ ms for atom numbers on the order of a few hundreds. For 
these systems our methods could be targeted at shorter times, as in 
the examples presented, providing an important improvement with 
respect to linear rampings. 

\begin{acknowledgments}
The authors thank M. W. Mitchell for a careful reading of the 
manuscript and useful suggestions. This work has been supported 
by FIS2011-24154, 2009-SGR1289, IT472-10, FIS2009-12773-C02-01, 
and the UPV/EHU under program UFI 11/55. B.~J.-D. is supported 
by the Ram\'on y Cajal program. 
\end{acknowledgments}

\appendix

\section{Shortcut equations for the Josephson junction with controllable 
linear coupling.}
\label{app:ex}
In this Appendix we shall transform the Schr\"odinger-like 
Eq.~(\ref{eq:tdse})  so that the invariant-based 
engineering technique for time-dependent harmonic oscillators 
developed in~\cite{muga1,muga2} may be applied. The structure of the 
Hamiltonian~(\ref{eq:parab}) is peculiar as it involves a time 
dependent (formal) mass factor in the kinetic-like term. The first 
step is to transform this Hamiltonian according to 
\beqa
\label{a1}
\eta\to\hbar,
\quad
4J(t)\to\frac{1}{m(t)},\quad
\frac{NU}{2}\to k \,,&&
\eeqa
to rewrite Eq.~(\ref{eq:parab}) as 
\begin{equation}
H = \frac{1}{2m(t)} {p}^2 + \frac{1}{2} k {z}^2, 
\label{eq:ms2}
\end{equation}
where $p=-i\hbar\partial_z$ is the ``momentum'' conjugate to 
$z$~\footnote{We shall use the symbol $p$ also for the momentum 
eigenvalues since the context makes its meaning clear.}. These 
and other transformations performed below are formal so that the 
dimensions do not necessarily correspond to the ones suggested 
by the symbols and terminology used. For example neither $p$, 
$m(t)$, or $z$ have dimensions of momentum, mass and length, respectively.  
 
Multiplying the time-dependent Schr\"odinger equation corresponding to 
Eq.~(\ref{eq:ms2}) from the left by momentum eigenstates $\langle p|$,
\begin{eqnarray}
i\hbar \partial_t \Psi(p,t)\! &=&\!
-\frac{\hbar^2 k}{2} \frac{\partial^2}{\partial p^2} \Psi(p,t)\! 
+\! \frac{p^2}{2m(t)} \Psi(p,t).
\label{eq:ms3}
\end{eqnarray}
Finally with the new mapping
\beqa\label{second}
k&\to&\frac{1}{m_x},
\\
\frac{1}{m(t)}&\to& m_x\omega_x^2(t),
\nonumber\\
p&\to& x \ ,
\nonumber
\eeqa
the Hamiltonian takes the standard time-dependent harmonic 
oscillator form 
\beq
\label{ordinary}
H=-\frac{\hbar^2}{2m_x}\frac{\partial^2}{\partial x^2}+\frac{1}{2}m_x\omega_x^2(t) x^2 \ .
\eeq
\begin{figure}[t]
\begin{center}
\includegraphics[width=62mm, angle=0]{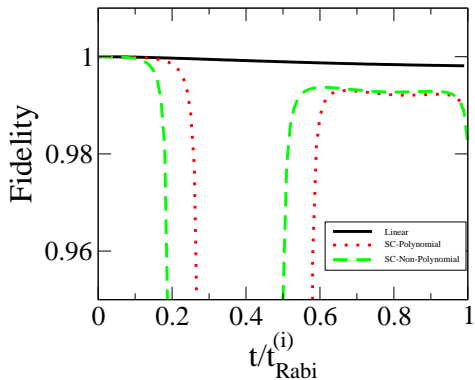}
\caption{Instantaneous fidelity as a function of time. 
$\gamma_i=0.1$ and $\gamma_f=1$ with $t_f=t_{\rm Rabi}^{(i)}$ 
for the shortcut and a linear path. }
\label{j3}
\end{center}
\end{figure}
%
Note that, thanks to the above transformations and basis change, 
the kinetic-like and potential-like terms in the 
Hamiltonian~(\ref{eq:parab}) have interchanged their roles so that 
the time dependence of the formal mass has become a time dependence 
of the formal frequency in Eq.~(\ref{ordinary}), whereas $m_x$ is 
constant. Fast dynamics between $t=0$ and $t_f$, from $\omega_x(0)$ 
to $\omega_x(t_f)$ without final excitations for this Hamiltonian 
may be inverse engineered by solving for $\omega_x(t)$ in the Ermakov 
equation~\cite{muga1,muga2}
\beq
\ddot{\rho}+\omega_x^2(t)\rho=\frac{\omega^2_{0}}{\rho^3}\ , 
\eeq
where $\omega_0$ is in principle an arbitrary constant, and  
$\rho(t)$ is a scale factor for the state that we may design, 
e.g. with a polynomial, so that it satisfies the boundary conditions 
\beqa
\rho(0)&=&\left(\frac{\omega_0}{\omega_x(0)}\right)^{1/2},\;
\rho(t_f)=\left(\frac{\omega_0}{\omega_x(t_f)}\right)^{1/2}\ ,
\nonumber\\
\dot{\rho}(0)&=&\dot{\rho}(t_f)=\ddot{\rho}(0)=\ddot{\rho}(t_f)=0. 
\eeqa
Defining $b=\hbar/\rho$, choosing $\omega_0=k\hbar$, and undoing 
the changes~(\ref{second}) and~(\ref{a1}) we rewrite the Ermakov 
equation as Eq.~(\ref{eq:erm}) and the boundary conditions 
become those in Eq.~(\ref{boco}). 

\section{Limitations of the method}
\label{app:bd}

As explained in the main text, when we have done the mapping 
between the results in Appendix~\ref{app:ex} and the Bose-Hubbard 
Hamiltonian, we have assumed $NU/2\gg J$ ($ \gamma \gg 1$). Thus 
our mapping should not be valid for small values of $\gamma$. 
In Fig.~\ref{j3} we show that this is indeed the case. We show 
the predictions for the fidelity of the shortcut protocol 
for $\gamma_i=0.1$ and $\gamma_f=1$ for $t_f=t_{\rm Rabi}^{(i)}$. 
In this special case one finds a better fidelity with the linear 
ramping than with the shortcut path.

\end{document}